\documentclass[11pt,a4paper]{article}

\usepackage{latexsym}
\usepackage{amssymb}
\usepackage{amsmath}
\usepackage{amsfonts}
\usepackage{mathrsfs}
\usepackage{cite}

\newcommand{\be}{\begin{equation}}
\newcommand{\ee}{\end{equation}}

\def\a{\alpha}
\def\b{\beta}

\def\tr{{\rm tr}}
\def\tr{{\rm tr}\,}

\def\cN{{\cal N}}
\def\cD{{\cal D}}

\def\bea{\begin{eqnarray}}
\def\eea{\end{eqnarray}}
\def\nn{\nonumber}

\def\cN{{\cal N}}

\def\f{\frac}

\def\tr{{\rm tr}\,}

\def\nn{\nonumber}

\def\d{\delta}

\def\sB{\stackrel{\frown}{\square}}

\def\eq{\eqref}

 \numberwithin{equation}{section}

\def\a{\alpha}
\def\h\a{\hat\a}
\def\b{\beta}
\def\h\b{\hat{\b}}

\def\tr{{\rm tr}}
\def\tr{{\rm tr}\,}

\def\cN{{\cal N}}
\def\cD{{\cal D}}

\def\bea{\begin{eqnarray}}
\def\eea{\end{eqnarray}}
\def\nn{\nonumber}

\def\cN{{\cal N}}

\def\f{\frac}

\def\tr{{\rm tr}\,}

\def\bea{\begin{eqnarray}}
\def\eea{\end{eqnarray}}
\def\nn{\nonumber}

\def\d{\delta}

\def\G{\Gamma}

\def\sB{\stackrel{\frown}{\Box}}

\begin{document}
\begin{titlepage}

\begin{center}
\vspace{1cm} {\Large\bf Quantum calculation of the
low-energy effective action\\
\vspace{0.2cm}

in $5D$, $\cN=2$ SYM theory }
\vspace{1.5cm}

 {\bf
 I.L. Buchbinder\footnote{joseph@tspu.edu.ru }$^{\,a,b}$,
 E.A. Ivanov\footnote{eivanov@theor.jinr.ru}$^{\,c}$,
 B.S. Merzlikin\footnote{merzlikin@tspu.edu.ru}$^{\,d,a}$,
 }
\vspace{0.4cm}

{\it $^a$ Department of Theoretical Physics, Tomsk State Pedagogical
University,\\ 634061, Tomsk,  Russia \\
 \vskip 0.1cm
 $^b$ National Research Tomsk State University, 634050, Tomsk, Russia \\
 \vskip 0.15cm
 $^c$ Bogoliubov Laboratory of Theoretical Physics, JINR, 141980 Dubna, Moscow region,
 Russia \\ \vskip 0.15cm
 $^d$ Tomsk State University of Control Systems and Radioelectronics, 634050 Tomsk, Russia
 }
\end{center}
\vspace{0.4cm}

\begin{abstract}
We consider $5D, \, \cN=2$ supersymmetric Yang-Mills (SYM)
theory in $5D, \,\cN=1$ harmonic superspace as a theory
of the interacting adjoint $5D, \,\cN=1$ gauge multiplet and hypermultiplet.
Using the background superfield method, we
compute the leading low-energy contribution to the one-loop
effective action. The result of quantum calculations precisely matches
the effective action derived earlier in {\tt arXiv:1812.07206} on the pure symmetry grounds.
\end{abstract}

\end{titlepage}

\setcounter{footnote}{0} \setcounter{page}{1}


\section{Introduction}

The study of maximally extended supersymmetric gauge theories in
dimensions larger then four is basically motivated by the relationships
of such theories to the low-energy string/brane dynamics (see e.g. \cite{L,BL}). In the present paper we consider the quantum
field aspects of $\cN=2$ SYM theory in five dimensions. This
theory bears an obvious interest because of its various connections with
$6D$, $\mathcal{N}=(2,0)$ superconformal field theory
compactified on a circle \cite{Douglas,LPS,LPS1} and also as a nice example
of applications of the localization technique \cite{loc1,loc2,loc3,loc4,loc5}.
Quantum effective action can be thought of as a universal tool of analyzing connections between
the low-energy effects in string theory and in quantum field theory.

The leading term of the low-energy effective action of $5D$, $\cN=2$ SYM theory
depending on all fields of $5D$, $\cN=2$ vector gauge multiplet was
constructed in ref. \cite{BIS}. This was accomplished by the
method similar to that employed in \cite{BuIv} for a similar calculation in $4D$, $\cN=4$ SYM
theory. The latter was formulated in $\cN=2$ harmonic superspace as a theory of $\cN=2$ vector
gauge multiplet coupled to the hypermultiplet in adjoint representation. Such
a theory, being manifestly $\cN=2$ supersymmetric, possesses an
additional hidden on-shell $\cN=2$ supersymmetry. As a result, it proves to  enjoy the total $\cN=4$ supersymmetry. It was shown that the
effective action depending on both the gauge  multiplet and the hypermultiplet
can be found in a closed form, starting from the known effective action in
the $\cN=2$ gauge multiplet sector and invoking the invariance under the hidden
$\cN=2$ supersymmetry. Such a purely symmetry-based analysis allowed to
determine the effective action up to a numerical coefficient. To specify the
coefficient, one should carry out the explicit quantum calculation. The latter
was performed in \cite{BIP}, where the result of  \cite{BuIv} was entirely confirmed
and the unknown overall coefficient was fixed.

In ref. \cite{BIS}, $5D$, $\cN=2$ SYM theory was formulated
in $5D, \,\cN=1$ harmonic superspace as a theory of interacting $\cN=1$ gauge multiplet and hypermultiplet in the adjoint
representation. The theory is manifestly $\cN=1$ supersymmetric and, in addition,  possesses an implicit
on-shell $\cN=1$ supersymmetry. Its effective action
in the $\cN=1$ gauge  multiplet sector was calculated some time ago in \cite{Pletnev}.
Like in the $4D, \cN=4$ case, the total $\cN=2$ supersymmetric effective action
of this theory was restored in \cite{BIS} through the completion of the $\cN=1$
gauge  multiplet  action by the proper hypermultiplet-dependent terms,
such that the full expression for the effective action respect the
additional implicit $\cN=1$ supersymmetry. The resulting effective
action can be written as an integral over the full $5D, \,\cN=1$  superspace \cite{BIS},
\bea
 S_{\mathrm{eff}}^{\mathcal{N}=2}= c_{0} \int d^{13}z  W\left [
 \ln W+ \frac{1}{2} H(Z) \right ], \label{Seff}
 \eea
where
\begin{equation}
H(Z) = 1+ 2\ln \frac{1 + \sqrt{1+2Z}}{2} +\frac{2}{3} \frac{1}{1 +
\sqrt{1+2Z}} -\frac{4}{3}\sqrt{1+2Z}\,, \qquad Z=\frac{Q^{+a}
Q^{-}_{a}}{W^{2}}\,. \label{H}
\end{equation}
Here $c_{0}$ is an arbitrary real numerical coefficient, $W$ is the
$\cN=1$ gauge superfield strength and $Q^{+a}, \,  Q^{-}_{a}$ are the
hypermultiplet superfields in the harmonic superspace formulation\footnote{Actually, the superfield Lagrangian in
\eq{Seff} does not depend on the harmonic variables on shell.}. Let us point out once more that the
result (\ref{Seff}), (\ref{H}) was obtained, based on the purely symmetry consideration.

The aim of the present paper is to evaluate the leading low-energy
effective action of $5D$, $\cN=2$ SYM by the explicit calculation
of the one-loop effective action in the quantum superfield perturbation theory.
We  perform the quantum superfield derivation of the action (\ref{Seff}), (\ref{H})
and specify the one-loop value of the coefficient $c_0$. To preserve the classical symmetries
in the quantum case, we make use of the background superfield method in
$5D, \cN=1$ harmonic superspace. It is a $5D$ version of the method developed
earlier in \cite{Kuz06,Pletnev} (see also \cite{KL}). Following the
approach of \cite{BIS}, we formulate $5D, \,\cN=2$
gauge multiplet as a collection of $\cN=1$ gauge multiplet
and the hypermultiplet,  both being in the adjoint representation of gauge
group. In the process of calculation we assume that the  background superfields
align in the Cartan subalgebra of $su(2)$ algebra and obey the
classical equations of motion. Also we restrict our consideration to
the background superfields slowly varying  in space-time, as this approximation  is sufficient
for finding the low-energy effective action. The expression for the effective action is obtained as an integral over the analytic
harmonic subspace. After passing to the full superspace, this expression
reproduces the effective action of ref. \cite{BIS},  with $c_0 = \frac1{48\pi^2}$.

The paper is organized as follows. Section 2 sketches the
formulation of $5D$, $\cN=2$ SYM theory in $\cN=1$
harmonic superspace. In section 3 we describe the manifestly gauge
covariant and $\cN=1$ supersymmetry-preserving procedure for calculating the
one-loop effective action. Section 4 is devoted to the
evaluation of the leading low-energy contribution to the one-loop
effective action. In the last section  we give a brief summary of the results obtained and
indicate possible future directions of the study.


\section{The model}

Throughout the paper we use the  notations and conventions of
\cite{BIS} and \cite{Pletnev}. We formulate $\mathcal{N}=2$ SYM
theory in $5D, \mathcal{N}=1$ harmonic superspace in terms of
the gauge superfield $V^{++}$ and the hypermultiplet
one $q^{+}_{a}\equiv (q^{+},-\bar{q}^{+})$, $a=1,2$, both being analytic. The classical
action of the theory is written as
 \bea
S&=& \frac{1}{2g^{2}} \sum _{n=2}^{\infty } \frac{(-i)^{n}}{n} \tr
\int d^{13}z du_{1} \ldots du_{n} \frac{V
^{++}(z,u_{1})V^{++}(z,u_{2})\ldots V^{++}(z,u_{n})}{(u^{+}_{1}
u^{+} _{2}) (u^{+}_{2} u^{+}_{3})\ldots (u^{+}_{n} u^{+}_{1})} \nn \\
&& - \frac{1}{2g^{2}} \tr \int d\zeta ^{(-4)} q^{+a}
{\mathcal{D}}^{++} q^{+}_a\,, \label{S0}
 \eea
where $g$ is a coupling constant of mass-dimension $-1/2$. We denote the full superspace integration measure as $d^{13}z = d^5 x
(\cD^-)^4(D^+)^4$ and the analytic subspace measure as $d\zeta ^{(-4)} = d^5 x
(\cD^-)^4 du$, where $du$ stands for the integration over harmonics. The powers
of the covariant derivatives are defined as
$(\cD^\pm)^4=-\f{1}{32}(\cD^\pm)^2(\cD^\pm)^2$, where
$(\cD^\pm)^2=\cD^{\pm \hat{\a}}\cD^{\pm}_{ \hat{\a}}$
$\hat{\a},\hat{\b}=1,2$. The covariant harmonic derivative
$\cD^{++}$ acts on the hypermultiplet according to the rule \cite{GIOS},
$\cD^{++} q^+_a = D^{++} q^+_a  + i[V^{++}, q^+_a]$. The action
\eq{S0} is invariant under the gauge transformation
 \bea
 \d V^{++}=-\cD^{++}\Lambda\,, \qquad \d q^+_a = - [q^+_a,\Lambda]\,,
 \label{gtr}
 \eea
with an analytic superfield gauge parameter $\Lambda =
\Lambda(\zeta,u)$.

The classical equation of motion associated with the action \eq{S0} read
 \begin{equation}
(\mathcal{D}^{+})^{2} W + i [q^{+a},q_{+a}]= 0\,, \qquad \cD^{++}
q^+_a =0\,, \label{DW}
 \end{equation}
where $W =\frac{i}{8} (\mathcal{D}^{+})^{2} V^{--}$ is the superfield
strength of the gauge multiplet. Here we introduced the non-analytic
superfield $V^{--}$ as a solution of the harmonic
zero-curvature condition \cite{GIOS}
 \begin{equation}
D^{++} V^{--} - D^{--} V^{++} + i[V^{++},V^{--}] =0\,. \label{zero}
 \end{equation}

The action \eq{S0} is formulated in $5D, \cN=1$
harmonic superspace and hence respects the manifest  off-shell $\cN=1$
supersymmetry. Since the hypermultiplet is in the adjoint representation of gauge group, like $V^{++}$,
the action \eq{S0} also exhibits invariance under an additional implicit $\mathcal{N}=1$
supersymmetry. One can check that the transformations
\begin{equation}
\delta q^{+}_{a} = -\frac{1}{2} (D^{+})^{4} [\epsilon _{a\hat{\alpha
}} \theta ^{-\hat{\alpha }} V^{--}]\,, \qquad \delta V^{++} =
\epsilon ^{a}_{\hat{\alpha }} \theta ^{+\hat{\alpha }} q ^{+}_{a}\,,
\label{hidden}
\end{equation}
where $\epsilon ^{a}_{\hat{\alpha }}$ is the relevant anticommuting
parameter, leave the action \eq{S0} invariant.

\section{One-loop effective action}
We construct the one-loop effective action for $\cN=2$ SYM theory with the ``microscopic'' action \eq{S0}
within the background superfield field formulation. The background superfield
method in $5D, \cN=1$ harmonic superspace
\cite{Pletnev} is a direct generalization of the $4D, \cN=2$ one \cite{BFM,non-ren,BK98,KM01} and it is based on the
background-quantum splitting of the initial superfields into the 'background'  ${\bf V}^{++}, {\bf Q}^+_a$
and the 'quantum' $v^{+}, q^+_a$ parts:
 \be
 V^{++}\rightarrow {\bf V}^{++} + gv^{++}\,, \qquad q^+_a \rightarrow
 {\bf Q}^+_a + g q^+_a\,.
  \ee

While quantizing the gauge theory with the action \eq{S0} by the background superfield technique, we as usual impose the gauge-fixing
conditions  on the quantum gauge superfield $v^{++}$ only. Then we
introduce the gauge-fixing action and the corresponding ghost action.
One of the main features of the background superfield method is  that the original infinitesimal gauge symmetry \eq{gtr}
is separated into the 'background' and 'quantum' transformations:
 \bea
 &&\d {\bf V}^{ ++} = -D^{++}\Lambda - i[{\bf V}^{ ++}, \Lambda], \qquad \d v^{++} =
 i[\lambda,v^{++}], \nn \\
 && \d {\bf Q}^+_a = -[Q^+_a, \Lambda], \qquad\qquad\qquad\quad\,\, \d q^+_a = 0\,.
 \eea
By construction, the effective action calculated loop by loop depends
only on the background superfields and hance is invariant under
the background gauge transformations.

As was said, in the framework of the background (super)field method, we should fix the gauge
with respect to the quantum gauge transformations. We choose the gauge-fixing
function as in $4D$ case \cite{non-ren,BK98}
 \be
 F^{(+4)}=D^{++} v^{++}.
 \ee
Under the quantum gauge group it transforms as follows
 \be
 \label{dF}
 \d F^{(+4)}=(\cD^{++}(\cD^{++}\lambda+i[v^{++},\lambda])).
 \ee
Then the action of the corresponding Faddeev-Popov ghosts ${\bf b}, {\bf c}$ is written as \cite{BFM}
 \be
 S_{FP}=\tr\int d\zeta^{(-4)}\ {\bf b}(\cD^{++})^2{\bf c}.
\label{FP}
 \ee

The harmonic superfield effective action for $5D$  gauge theories
is constructed in the same way as in $4D$, $\cN=2$
\cite{BK98} and $6D$, $\cN=(1,0)$ \cite{BIMS-b} cases. For $5D$ supersymmetric gauge theories the background superfield
method was developed in refs. \cite{Kuz06} and \cite{Pletnev}. The one-loop quantum correction to the effective action
$\G^{(1)}[{\bf V}^{++},{\bf Q}^+]$ is defined by the functional
integral over quantum fields $v^{++}, q^+_a$ and ghosts fields as
 \bea
 \label{def}
 e^{i\G^{(1)}}= {\rm Det}_{(4,0)}^{1/2}{\boldsymbol\sB}\int \cD v^{++}\cD q^{+} \cD {\bf b}
\cD{\bf c}\cD\varphi
 \,\,e^{i S^{(2)}_{\rm quant}[v^{++}, q^+_a, {\bf b}, {\bf c}, \varphi, {\bf V}^{++}, {\bf Q}^+]},
 \eea
where the bilinear in quantum superfields part of the quantum action is
 \be
 \label{defq}
 S^{(2)}_{\rm quant}= S^{(2)}_{0} + S_{\rm gf} + S_{FP} + S_{NK},
 \ee
and we introduced the background-dependent operator ${\boldsymbol
\sB} = \f12 (D^{+})^4({\boldsymbol \cD}^{--})^2$. The definition of  the functional determinant 
${\rm Det}_{(4,0)}{\boldsymbol\sB}$ is given in ref. \cite{BK98}. 
On a space of analytical superfields the operator ${\boldsymbol \sB}$ is reduced to \cite{Kuz06}
 \bea
{\boldsymbol\sB} = {\boldsymbol
\cD}^{\hat{a}}{\boldsymbol\cD}_{\hat{a}} +(D^{+\hat{\a}}{\bf W})
{\boldsymbol\cD}^-_{\hat{\a}} - \f14(D^{+\hat{\a}}D^+_{\hat{\a}}{\bf
W}){\boldsymbol\cD}^{--} +
\f14(D^{+\hat{\a}}{\boldsymbol\cD}^-_{\hat{\a}}{\bf W}) - {\bf
W}^2\,.
 \eea
Here, all 'bold' symbols involve only the background gauge multiplet. For instance, the covariant space-time derivative is written
through the background gauge connection as ${\boldsymbol \cD_{\hat a}
= \partial_{\hat{a}}}-i{\bf A}_{\hat{a}}, \,\,{\hat a} = 0,..,4\,$.

The quadratic action \eq{defq} includes the Faddeev-Popov ghost action \eq{FP}, in which
the harmonic covariant derivative depends on
the background superfield ${\bf V}^{++}$, and the action for
Nielsen-Kallosh ghost $\varphi$
 \bea
 S_{NK} = \f12 \tr \int d\zeta^{(-4)} \varphi ({\boldsymbol
 \cD}^{++})^2 \varphi\,.
 \eea
The action \eq{defq} also contains the sum of the
quadratic part of the classical action $S_0$ and the gauge-fixing action
$S_{\rm gf}$
 \bea
 S^{(2)}_0+S_{\rm gf} &=& -\frac{1}{2}\tr\int d\zeta^{(-4)}\, v^{++}{\boldsymbol\sB} v^{++}
  -\f12 \tr\int d\zeta^{(-4)}\,{q}^{+ a}{\boldsymbol\cD}^{++} q^{+}_{a} \nn \\
 &&- \f{i}{2}\tr \int d\zeta^{(-4)}\Big\{
  {\bf Q}^{+ a}[ v^{++}, q^{+}_{ a}] + {q}^{+ a}[v^{++}, {\bf
  Q}^{+}_{a}]\Big\}\,.
  \label{S2}
 \eea

The action \eq{S2} involves terms which mix the quantum gauge
multiplet $v^{++}$ and the quantum hypermultiplet $q^{+}_a$. These terms can be
eliminated in $R_{\xi}$ gauge (see, e.g., \cite{BBM} for an example
of application of  the $R_{\xi}$ gauge in $6D, \cN=(1,1)$ SYM theory). In this case the action for the Faddeev-Popov
ghosts would depend on both the background gauge multiplet and
hypermultiplet and involve inverse powers of the operator $\sB\,$. Instead of
imposing $R_\xi$ gauge, we use a special change of quantum hypermultiplet
\cite{BIMS-b} in the functional integral \eq{def}
 \bea
 \label{replac}
 q^{+}_{a}(1)= h^{+}_{a}(1) - i\int d \zeta^{(-4)}_2\,
 G^{(1,1)}(1|2)_a{}^b [v^{++}(2), {\bf Q}^{+}_{b}(2)]\,,
 \eea
with $h^{+}_{a}$ being a set of new independent quantum superfields.
The change \eq{replac} leads to the cancelation of mixed terms in the
action \eq{S2}. The Jacobian of the change (\ref{replac}) equals one and so it does not affect the integration measure in
\eq{def}. After changing  the variables as in \eq{replac}, he action \eq{S2} acquires the form
 \bea
 S_0^{(2)}+S_{\rm gf} &=& \frac{1}{2} \tr\int d\zeta_1^{(-4)}\,d\zeta_2^{(-4)}\,
 v_1^{++}\Big\{{\boldsymbol\sB} \d^{(3,1)}_A(1|2)
 - 2\,{\bf Q}^{+ a}(1) G^{(1,1)}(1|2){\bf Q}^{+}_{a}(2)\Big\}v_2^{++} \nn\\
 &&-\f12 \tr\int d\zeta^{(-4)}\, h^{+a}{\boldsymbol \cD}^{++} h^+_a\,.
 \label{S2v}
 \eea
The Green function appearing in \eq{replac} and \eq{S2v},
$G^{(1,1)}(\zeta_1,u_1|\zeta_2,u_2)_a{}^b = i\langle0| {\rm
T}{q}^{+}_{a}(\zeta_1,u_1) {q}^{+\,b}(\zeta_2,u_2)|0\rangle\,,$ is the
background-dependent superfield hypermultiplet Green function in the
$\tau$-frame. It is analytic with respect to its both arguments
and satisfies the equation
 \bea
 \label{eqG}
 {\boldsymbol \cD}_1^{++}G^{(1,1)}(1|2)_a{}^b &=&\delta_a{}^b \d_{\cal A}^{(3,1)}(1|2)\,.
 \eea
In the $\tau$-frame the Green function can be written as $G^{(1,1)}(1|2)_a{}^b = \d_a{}^b G^{(1,1)}(1|2)$, where
 \bea
 G^{(1,1)}(1|2) &=&  \f{(D^+_1)^4(D^+_2)^4}{{\boldsymbol\sB}_1}
 \f{\d^{14}(z_1-z_2)}{(u^+_1u^+_2)^3}\,,
 \label{GREEN}
  \eea
and  $\d_A^{(3,1)}(1|2)$ is a covariantly-analytic delta-function
\cite{GIOS}.

In the action \eq{def} the background superfields ${\bf V^{++}} $
and ${\bf Q}^+_a$ are analytic but unconstrained otherwise. The
gauge group of the theory \eq{S0} is assumed to be $SU(2)$. For further consideration,
we will also assume that the background
fields ${\bf V}^{++}$ and  ${\bf Q}^+_a$ align in the Cartan
subalgebra of $su(2)$
 \bea
 \label{background}
 {\bf V}^{++} = V^{++}(\zeta,u) H\,, \qquad {\bf Q}^+_a =  Q^+_a(\zeta,u)\, H\,,
 \eea
where $H=\f12 \sigma_3$ and $\sigma_3$ is Pauli matrix. The components of the background superfields associated
with the $E_\pm$ generators ($[E_+,E_-]=2H$ and $[H,E_\pm]=\pm E_\pm$) are assumed to vanish. Our choice of the
background corresponds to the spontaneous symmetry breaking $SU(2)
\rightarrow U(1)$. We denote the non-zero components of background
superfield ${\bf V}^{++}$ in \eq{background} by the same letter
$V^{++}$ as in the classical action \eq{S0}, with the hope that this will not result in a confusion.
The same remark refers to the abelian superfield strength $W$ constructed out of $V^{++}$.

We assume that the background superfields ${\bf V}^{++}$ and ${\bf
Q}^+_a$ satisfy the classical equations of motion \eq{DW}. The conditions
\eq{background} then imply free equations of motion for the superfields $V^{++}$ and $Q^+_a\,$,
\bea
 D^{+\hat{\a}} D^+_{\hat \a} W =0\, \qquad D^{++} Q^+_a = 0\,.
 \eea
We also consider the case of the slowly varying background gauge superfield
strength and hypermultiplet
 \bea
 \partial_{\hat a} W \simeq 0\, \qquad
 \partial_{\hat a} Q^+_a \simeq 0\,.
 \eea
With our choice of the  background superfields as described
above, it is convenient to rewrite the implicit supersymmetry transformations (\ref{hidden}) in terms
of the gauge superfield strength \cite{BIS}. They are
 \begin{equation}
 \delta Q^{+}_{a} = \frac{i}{2} \epsilon _{a}^{\hat{\alpha }}
 (D^{+}_{\hat{\alpha }} W)\,, \qquad
 \delta W = -\frac{i}{4}
 \epsilon^{a}_{\hat{\alpha }} D^{- \hat{\alpha }} Q^{+}_{a}\,. \label{hidden1}
 \end{equation}

The further strategy is as follows. We substitute \eq{background} in
the action \eq{S2v} and in  the actions for the ghosts superfields $S_{FP}$ and
$S_{NK}$. As the next step, we integrate over quantum superfields $v^{++}$
and $h^+_a$ in the functional integral \eq{def}. As in
$4D$ \cite{BK98} and $6D$ \cite{BIMS-b} cases
the contributions of the ghost superfields exactly cancel the
contribution of the quantum hypermultiplet. Thus we are left with  the
difference between the contribution from the quantum gauge multiplet $v^{++}$
and the contribution from the additional determinant ${\rm Det}_{(4,0)}^{1/2}{\boldsymbol\sB}$ in \eq{def}. The presence of this
determinant is necessary for eliminating the contributions from the longitudinal
component of the superfield $v^{++}$, in full analogy with $4D$ and $6D$
cases (see \cite{BIP}, \cite{KM01} and \cite{BIM-19}). Finally, for the one-loop contribution $\Gamma^{(1)}$ to the effective
action we obtain the expression
 \bea
\Gamma^{(1)} = i{\rm Tr}_{\rm T} \ln\Big(
{\cD}^{\hat{a}}{\cD}_{\hat{a}} +(D^{+\hat{\a}}W) D^-_{\hat{\a}} -
{W}^2 - 2\,{Q}^{+ a} G^{(1,1)}{Q}^{+}_{a}\Big), \label{1loop}
 \eea
where we executed the trace over matrix indices. The functional
trace in \eq{1loop} is defined as a trace over the transversal
component of the superfield $v^{++}$
 \bea
\Gamma^{(1)} = i \int d\zeta^{(-4)} \ln\Big(
{\cD}^{\hat{a}}{\cD}_{\hat{a}} +(D^{+\hat{\a}}W) D^-_{\hat{\a}} -
{W}^2 - 2\,{Q}^{+ a} G^{(1,1)}{Q}^{+}_{a}\Big) \Pi^{(2,2)}_{\rm T}
(1|2)\Big|_{2\to 1}\,,
 \eea
where the projector $\Pi^{(2,2)}_{\rm T} (1|2)$ is analytic in both
arguments and is defined as \cite{KM01}
 \bea
\Pi^{(2,2)}_{\rm T}(1|2) =\d^{(2,2)}_{\cal
A}(1|2)-\cD^{++}_1\cD^{++}_2\f{(D^+_1)^4(D^+_2)^4}{\sB_1}
\d^{13}(z_1-z_2)\f{(u^-_1u^-_2)}{(u^+_1u^+_2)^3},
 \eea
with $\d^{(2,2)}_{\cal A}(1|2)$ being an analytic delta-function
\cite{GIOS}. For our calculation, we do not need to know the explicit form of
the projector $\Pi^{(2,2)}_{\rm T}$, which for the $5D$ case was
found in \cite{Pletnev}. All what we need is the
expression in the limit of coincident harmonic arguments,
$u_2\to u_1$, in the case of slowly varying on-shell background
superfields. The latter condition implies the simplest form  for the projector $\Pi^{(2,2)}_{\rm T}\,$,
\bea
\label{pro} \Pi^{(2,2)}_{\rm T}(1|2)\big|_{u_2\to
u_1}=(D_1^+)^4\d^{13}(z_1-z_2).
 \eea

Thus we finally arrive at the following expression for the one-loop contribution
$\Gamma^{(1)}$ \eq{1loop}:
 \bea
\Gamma^{(1)} = i \int d\zeta^{(-4)} \ln\Big(
{\cD}^{\hat{a}}{\cD}_{\hat{a}} +(D^{+\hat{\a}}W) D^-_{\hat{\a}} -
{W}^2 - 2\,{Q}^{+a}G^{(1,1)}{Q}^{+}_{a}\Big)
(D_1^+)^4\d^{13}(z_1-z_2) \Big|_{2\to 1}\,. \label{1loop2}
  \eea
It is the starting point for the evaluation of the leading
low-energy contribution to the effective action in the model under
consideration.

\section{Leading low-energy contribution}

Here we demonstrate how the exact expression for the leading
low-energy contribution to effective action derived  in \cite{BIS}
can be recovered from the one-loop effective action \eq{1loop2}.

First of all we have to note that the effective action \eq{1loop2}
contains the non-local contribution ${Q}^{+ a}(1)
G^{(1,1)}(1|2){Q}^{+}_{a}(2)$ and so one should extract the local part
from it. One can use the following identity \cite{Pletnev}
 \bea
(D^+_1)^4(D^+_2)^4\f{1}{(u^+_1u^+_2)^3} =
(D^+_1)^4\Big\{(\cD_1^-)^4(u^+_1u^+_2)-\f14(u^-_1u^+_2)\Delta^{--}_1-
\f{(u^-_1u^+_2)^2}{(u^+_1u^+_2)}\sB_1\Big\}, \label{id}
 \eea
where $
\Delta^{--}=i\cD^{\hat{\a}\hat{\b}}\cD^-_{\hat\a}\cD^-_{\hat\b}
 +W(\cD^-)^2+4(\cD^{-\hat{\a}}W)\cD^-_{\hat{\a}}$. An analog of this identity for  $4D, \cN=2$ supersymmetric gauge
theory  was originally derived in \cite{KM01}. Then we use the decomposition $Q^+_a(2) =
(u^+_1 u^+_2)Q_a^-(1) - (u^-_1 u^+_2)Q_a^+(1)$ and eq. \eq{id} to
properly transform the Green function $G^{(1,1)}(1|2)$ \eq{GREEN}\footnote{See
the detailed analysis of the similar contribution in $6D$,
$\cN=(1,1)$ SYM theory in ref. \cite{BIM-19}.}. We have
 \bea
 {Q}^{+ a}(1) G^{(1,1)}(1|2){Q}^{+}_{a}(2) = Q^{+a}Q^-_a (u^-_1 u^+_2)^2
 \delta^{13}(z_1-z_2) + \ldots,
 \eea
where dots stand for terms proportional to
$(u^+_1u^+_2)$ and so vanishing in the $u_2\to u_1$ limit.

Thus we obtain for the one-loop contribution \eq{1loop2}
 \bea
\Gamma^{(1)} = i \int d\zeta^{(-4)} \ln\Big(
{\cD}^{\hat{a}}{\cD}_{\hat{a}} +(D^{+\hat{\a}}W) D^-_{\hat{\a}} -
{W}^2 - 2{Q}^{+a}{Q}^{-}_{a}\Big) (D_1^+)^4\d^{13}(z_1-z_2)
\Big|_{2\to 1}\,. \label{1loop3}
  \eea
In order to evaluate the leading low-energy contribution to
the effective action we need to calculate the functional trace in
\eq{1loop3} in the coincident-point limit. First we calculate the
$\theta^{\pm}_2 \to \theta^\pm_1$ limit using the presence of
Grassmann delta-functions in \eq{1loop3}. In the full
superspace delta-function
 \be
\d^{13}(z_1-z_2) =
\d^5(x_1-x_2)\d^4(\theta^+_1-\theta^+_2)\d^4(\theta^-_1-\theta^-_2)
 \ee
the operator $(D^+)^4$ annihilate one of the Grassmann delta-functions according to the rule
 \bea
 (D^+)^4 \delta^4(\theta^-_1-\theta^-_1) = -2\,.
 \eea
In order to remove the remaining delta-function
$\d^4(\theta^-_1-\theta^-_2)$, we need to collect the fourth power of the
derivative $D^-_{\hat \a}$. To this end, we expand the logarithm in \eq{1loop3} in
the power series, up to the forth power of $(D^{+\hat{\a}}W)
D^-_{\hat{\a}}$:
 \bea
 \Gamma^{(1)} = \f{i}{2}\int d\zeta^{(-4)}
 \f{(D^{+\hat{\a}}W D^+_{\hat{\a}}W)^2}
 {(\partial^{\hat{a}}\partial_{\hat{a}} - W^2 -2Q^{+ a}Q^-_a)^4}
 \d^5(x_1-x_2) +\dots,
 \label{1loop4}
 \eea
where dots mean all contribution with the derivatives of
hypermultiplet, $D^{-}_{\hat{\a}}Q^{+a}$, which can in principle be
evaluated explicitly. In what follows we omit all such
contributions, assuming that they can be reconstructed by using the
analyticity condition for the integrand in \eq{1loop4} and the
implicit $\cN=1$ supersymmetry \eq{hidden1}.

Then we pass to the momentum representation for the space-time
delta-function and calculate the momentum integral
 \bea
 \int \f{d^5 p}{(2\pi)^5}\f1{(p^2+M^2)^4} = \f{i}{6 (8\pi)^2}\f1{M^3}\,.
 \eea
After that we obtain for the $\Gamma^{(1)}$ \eq{1loop4} the following expression
 \bea
\Gamma^{(1)} =-\f{1}{12(8\pi)^2} \int d\zeta^{(-4)}
\f{(D^{+\hat{\a}}W D^+_{\hat{\a}}W)^2}{(W^2 +2Q^{+ a}Q^-_a)^{3/2}} + \dots,
 \label{1loop5}
 \eea
where dots as in (\ref{1loop4}) mean terms with derivatives of the hypermultiplet. The expression \eq{1loop5}
is the leading low-energy contribution to the effective action of
$5D, \cN=5$ SYM theory. It is written as an integral over the
analytic subspace.  In the paper \cite{BIS} the effective action of
$\cN=2$ SYM theory was obtained as a hypermultiplet completion of
the leading $W \ln W$-term in the $\mathcal{N}=1$ SYM low-energy
effective action and it was written as an integral over the whole
superspace. This effective action was evaluated in the form
(\ref{Seff}), up to an overall constant $c_0$.

Let us demonstrate that,  passing to the full superspace in \eq{1loop5},  one can reproduce the expression  \eq{Seff}.
To this end, we first expand the function $H(Z)$ in the expression
\eq{Seff} in the power series
 \begin{equation}
S_{\mathrm{eff}}^{\mathcal{N}=2}= c_0\int d^{13}z\left [W \ln W +
\sum _{n=1}^{\infty }\frac{(-1)^{n}(2n-2)!}{n!(n+1)!
2^{n}}\frac{(Q^{+a}Q^{-} _{a})^{n}}{W^{2n-1}} \right]\,.
\label{Gamma}
\end{equation}
Then we decompose the factor $(W^2 +2Q^{+ a}Q^-_a)^{-3/2}$ in \eq{1loop5} as
 \bea
\Gamma^{(1)}=-\f{1}{12(8\pi)^2} \int d\zeta^{(-4)} (D^{+\hat{\a}}W
D^+_{\hat{\a}}W)^2\Big(\f{1}{W^3}+ \sum_{n=1}^{\infty}\f{(-1)^n
2^{n}\Gamma[n+\f32]}{\Gamma[\f32]\,\Gamma[n+1]}\f{(Q^{+a}Q^{-}
_{a})^{n}}{W^{2n+3}}\Big).\label{1loop6}
 \eea
After this we pass to the full superspace by restoring $(D^+)^4$ in all terms of the series by the rules
 \bea
(D^+)^4 W\ln W &=& -\f{1}{16}\f{(D^{+\hat{\a}}W
D^+_{\hat{\a}}W)^2}{W^3}\,, \nn \\
(D^+)^4 \f{1}{W^{2n-1}} &=& -\f{1}{8}
n(n+1)(2n+1)(2n-1)\f{(D^{+\hat{\a}}W
D^+_{\hat{\a}}W)^2}{W^{2n+3}}\,, \label{coeff}
 \eea
keeping in mind the on-shell condition for the background gauge
field strength $W$ and omitting all terms with derivatives of the
hypermultiplet. One can show that, after using \eq{coeff} in
\eq{1loop6} and employing the property,
$$\Gamma[n+\tfrac12] = \f{2n! \sqrt{\pi}}{4^n n!},$$ the second term in \eq{1loop6} immediately takes
the same form as in \eq{Gamma}. Indeed, it is straightforward
to check that
 \bea
 (D^+)^4 \Big(W\ln W + \tfrac12 W H\big(\tfrac{Q^{+a}Q^-_a}{W^2}\big)\Big)
  =-\f{1}{16} \f{(D^{+\hat{\a}}W D^+_{\hat{\a}}W)^2}{(W^2 +2Q^{+
  a}Q^-_a)^{3/2}} + \dots,
 \eea
where dots denote terms with spinor derivatives of
the hypermultiplet. Thus  for the leading term in the low-energy effective action we obtain the expression
 \bea
 \Gamma^{(1)} = \f{1}{48\pi^2}\int d^{13}z W\left [
 \ln W+ \frac{1}{2} H\Big(\tfrac{Q^{+a}Q^-_a}{W^2}\Big) \right]\,,
 \eea
where the function $H(Z)$ was defined in \eq{H}.

We observe the complete agreement of the method based on the symmetry considerations
with the direct quantum computations. The latter also yield the precise value for the coefficient $c_0$.

\section{Summary}
We have studied the  problem of computing the leading contribution to the one-loop
low-energy effective action of $5D$, $\mathcal{N}=2$ SYM theory in the $5D, \, \cN=1$ harmonic superspace formulation. The effective
action was constructed in the framework of the background field
method which  allow to preserve the manifest gauge invariance and
$5D, \, \cN=1$ supersymmetry at all stages of calculations.  The effective action derived in this way
depends on all fields of $5D$, $\mathcal{N}=2$ gauge multiplet  and is completely $5D$,
$\mathcal{N}=2$ supersymmetric. We have shown that the superfield
quantum considerations yield the same leading contribution to
one-loop low-energy effective action as the analysis carried out in ref. \cite{BIS}
on the purely symmetric grounds.

The results obtained here can be further generalized at
least in two directions. First, it would be interesting  to find out
the explicit forms of the next-to-leading corrections to the
effective action (\ref{Seff}). Second, it is tempting to
study the quantum aspects of the twisted $5D,\ \mathcal{N}=2$ SYM
theory \cite{Witten11,Haydys,Zabzine}, using similar techniques.

\section*{Acknowledgments}

\noindent This research was supported in part by RFBR grant, project
No. 18-02-01046, and Russian Ministry of Education and Science grant,
project No. 3.1386.2017.  The work of B.S.M. was supported in part
by the Russian Federation President grant, project
MK-1649.2019.2.



\begin{thebibliography}{66}
   \setlength{\itemsep}{0mm}

%
\bibitem{L} N. Lambert, {\it M-theory and maximally supersymmetric gauge theories},
Ann. Rev. Nucl. Part. Sci. {\bf 62} (2012) 285, {\tt arXiv:1203.4244
[hep-th]}.
%
\bibitem{BL} J. Bagger, N. Lambert, S. Mukhi, C. Papageorgakis,
{\it Multiple membranes in M-theory}, Phys. Rept. {\bf 527} (2013)
1, {\tt arXiv:1203.3546 [hep-th]}.
%
\bibitem{Douglas} M.~R.~Douglas, {\it On D=5 super Yang-Mills theory and (2,0) theory},
JHEP {\bf 1102} (2011) 011, {\tt arXiv:1012.2880 [hep-th]}.

\bibitem{LPS} N.~Lambert, C.~Papageorgakis and M.~Schmidt-Sommerfeld, {\it M5-Branes,
D4-Branes and Quantum 5D
super-Yang-Mills}, JHEP {\bf 1101} (2011) 083, {\tt arXiv:1012.2882
[hep-th]}.

\bibitem{LPS1} N.~Lambert, C.~Papageorgakis and M.~Schmidt-Sommerfeld,
{\it Deconstructing (2,0) Proposals}, Phys. Rev. D {\bf 88} (2013)
026007, {\tt arXiv:1212.3337 [hep-th]}.

\bibitem{loc1} K.~Hosomichi, R.-K.~Seong and S. Terashima,
{\it Supersymmetric gauge theories on the five-sphere}, Nucl. Phys.
B {\bf 865} (2012) 376, {\tt arXiv:1203.0371 [hep-th]}.

\bibitem{loc2} J.~Kallen, J.~Qiu and M. Zabzine,
{\it The perturbative partition function of supersymmetric 5D
Yang-Mills theory with matter on the five-sphere}, JHEP {\bf 08}
(2012) 157, {\tt arXiv:1206.6008 [hep-th]}.

\bibitem{loc3} J.~Kallen, J.~A.~Minahan, A.~Nedelin and M.~Zabzine,
{\it $N^3$-behavior from 5D Yang-Mills theory}, JHEP {\bf 1210}
(2012) 184, {\tt arXiv:1207.3763 [hep-th]}.

\bibitem{loc4} H.-C.~Kim and S.~Kim, {\it M5-branes from gauge theories on the
5-sphere}, JHEP {\bf 1305} (2013) 144, {\tt arXiv:1206.6339
[hep-th]}.

\bibitem{loc5} Y.~Imamura, {\it Perturbative partition function for a squashed
$S^5$}, PTEP {\bf 2013} (2013) 073B01, {\tt arXiv:1210.6308
[hep-th]}.

\bibitem{BIS} I.L. Buchbinder, E.A. Ivanov, I.B. Samsonov, {\it Low-energy effective
action in 5D, N=2 supersymmetric gauge theory}, Nucl. Phys. {\bf B}
940 (2019) 54, {\tt arXiv:1812.07206 [hep-th]}.

\bibitem{BuIv} I.~L.~Buchbinder and E.~A.~Ivanov,
{\it Complete N=4 structure of low-energy effective action in N=4
superYang-Mills theories}, Phys. Lett. B {\bf 524} (2002) 208, {\tt
hep-th/0111062}.

\bibitem{BIP} I.~L.~Buchbinder, E.~A.~Ivanov and A.~Yu.~Petrov,
{\it Complete low-energy effective action in N=4 SYM: A direct N=2
supergraph calculation}, Nucl. Phys. B {\bf 653} (2003) 64, {\tt
hep-th/0210241}.

\bibitem{Pletnev}
  I.~L.~Buchbinder and N.~G.~Pletnev,
  {\it Effective actions in $ \mathcal{N}=1$,
  D5 supersymmetric gauge theories: harmonic superspace approach},
  JHEP {\bf 1511} (2015) 130,
  {\tt arXiv:1510.02563 [hep-th]}.

\bibitem{Kuz06} S.~M.~Kuzenko, {\it Five-dimensional supersymmetric Chern-Simons
action as a hypermultiplet quantum correction}, Phys. Lett. B {\bf
644} (2007) 88, {\tt hep-th/0609078}.

\bibitem{KL} S.~M.~Kuzenko, W.~D.~Linch, III,
{\it On five-dimensional superspaces}, JHEP {\bf 0602} (2006) 038,
{\tt hep-th/0507176}.


\bibitem{GIOS}
A. S. Galperin, E. A. Ivanov, V. I. Ogievetsky, E. S. Sokatchev,
{\it Harmonic Superspace}, Cambridge University Press, Cambridge,
2001, 306 p.
%
\bibitem{non-ren}
  I.~L.~Buchbinder, S.~M.~Kuzenko and B.~A.~Ovrut,
  {\it On the D=4, N=2 nonrenormalization theorem},
  Phys.\ Lett.\ B {\bf 433} (1998) 335,
  {\tt hep-th/9710142}.

\bibitem{BFM}
  I.~L.~Buchbinder, E.~I.~Buchbinder, S.~M.~Kuzenko and B.~A.~Ovrut,
  {\it The background field method for N=2 super
  Yang-Mills theories in harmonic superspace},
  Phys.\ Lett.\ B {\bf 417} (1998) 61, {\tt hep-th/9704214}.

\bibitem{BK98}
  I.~L.~Buchbinder and S.~M.~Kuzenko,
  {\it Comments on the background field method in harmonic superspace:
  Nonholomorphic corrections in N=4 SYM},
  Mod.\ Phys.\ Lett.\ A {\bf 13} (1998) 1623,
  {\tt hep-th/9804168}.


\bibitem{KM01}
  S.~M.~Kuzenko and I.~N.~McArthur,
  {\it Effective action of N=4 super Yang-Mills: N=2 superspace approach},
  Phys.\ Lett.\ B {\bf 506} (2001) 140,
  {\tt hep-th/0101127}.

%
\bibitem{BIMS-b}
I.L. Buchbinder, E.A. Ivanov, B.S. Merzlikin, K.V. Stepanyantz, {\it
One-loop divergences in 6D, N=(1,0) SYM theory}, JHEP {\bf 1701}
(2017) 128, {\tt arXiv:1612.03190 [hep-th]}.


%
\bibitem{BBM}I.~L.~Buchbinder, A.S. Budekhina, B.~S.~Merzlikin,
{\it On the component structure of one-loop effective actions in
$6D$, $\cN=(1,0)$ and $\cN=(1,1)$ supersymmetric gauge theories},
{\tt arXiv:1909.10789 [hep-th]}.

\bibitem{BIM-19} I.L. Buchbinder, E.A. Ivanov, B.S. Merzlikin, {\it
Low-energy 6D, N=(1,1) SYM effective action beyond the leading
approximation}, {\tt arXiv:1912.02634 [hep-th]}.



\bibitem{Witten11} E. Witten, {\it Fivebranes and Knots}, {\tt arXiv:1101.3216 [hep-th]}.

\bibitem{Haydys} A. Haydys, {\it Fukaya-Seidel category and gauge theory},
J. Sympl. Geom. {\bf 13} (2015) 151, {\tt arXiv:1010.2353
[math.SG].}

\bibitem{Zabzine} J. Qiu and M. Zabzine, {\it
On twisted $N=2$ $5D$ super Yang-Mills theory}, Lett. Math. Phys.
{\bf 106} (2016) 1, {\tt arXiv:1409.1058 [hep-th]}.


\end{thebibliography}
\end{document}